\documentclass[fleqn,usenatbib]{mnras}

\usepackage[T1]{fontenc}

\DeclareRobustCommand{\VAN}[3]{#2}
\let\VANthebibliography\thebibliography
\def\thebibliography{\DeclareRobustCommand{\VAN}[3]{##3}\VANthebibliography}

\usepackage{graphicx}	
\usepackage[normalem]{ulem}

\usepackage{amsmath,amsfonts,amssymb}

\usepackage{caption}
\usepackage{subcaption}

\title[Accretion in triples with discs]{Accretion rates in hierarchical triple systems with discs}

\author[S. Ceppi et al.]{\parbox{\textwidth}{
Simone Ceppi,$^{1}$\thanks{corresponding author: simone.ceppi@unimi.it}
Nicol\'as Cuello,$^{2}$
Giuseppe Lodato,$^{1}$
Cathie Clarke,$^{3}$
Claudia Toci,$^{1}$
Daniel J. Price$^{4}$}
\\
$^{1}$Dipartimento di Fisica, Universit\`a Degli Studi di Milano, Via Celoria, 16, Milano, 20133, Italy.\\
$^{2}$Univ. Grenoble Alpes, CNRS, IPAG, F-38000 Grenoble, France.\\
$^{3}$Institute of Astronomy, University of Cambridge, Madingley Road, CB3 0HA, Cambridge, UK\\
$^{4}$School of Physics and Astronomy, Monash University, Clayton, VIC 3800, Australia}

\date{Accepted XXX. Received YYY; in original form ZZZ}

\pubyear{2021}

\begin{document}
\label{firstpage}
\pagerange{\pageref{firstpage}--\pageref{lastpage}}
\maketitle

\begin{abstract}
Young multiple systems accrete most of their final mass in the first few Myr of their lifetime, during the protostellar and protoplanetary phases. Previous studies showed that in binary systems the majority of the accreted mass falls onto the lighter star, thus evolving to mass equalisation. However, young stellar systems often comprise more than two stars, which are expected to be in hierarchical configurations. Despite its astrophysical relevance, differential accretion in hierarchical systems remains to be understood. In this work, we investigate whether the accretion trends expected in binaries are valid for higher order multiples. We performed a set of 3D Smoothed Particle Hydrodynamics simulations of binaries and of hierarchical triples (HTs) embedded in an accretion disc, with the code {\sc Phantom}. We identify for the first time accretion trends in HTs and their deviations compared to binaries. These deviations, due to the interaction of the small binary with the infalling material from the circum-triple disc, can be described with a semi-analytical prescription. Generally, the smaller binary of a HT accretes more mass than a single star of the same mass as the smaller binary. We found that in a HT, if the small binary is heavier than the third body, the standard differential accretion scenario (whereby the secondary accretes more of the mass) is hampered. Reciprocally, if the small binary is lighter than the third body, the standard differential accretion scenario is enhanced. The peculiar differential accretion mechanism we find in HTs is expected to affect their mass ratio distribution. 
\end{abstract}




\section{Introduction}
\label{sec:intro}
Surveys of star forming regions indicate that multiple stellar systems are common in young populations \citep{Reipurth+14,Duchene&Kraus13}. Among Class 0 and Class I stars (younger than 1 Myr) the multiplicity fraction ranges between 40\% and 70\% \citep{Connelley+08,Chen+13}, while in evolved populations is around 20\% \citep{Duquennoy&Mayor91}. In addition, molecular cloud simulations show that protostars are likely to form as part of multiple stellar systems and that their surrounding discs experience dramatic dynamical interactions with neighbour stars \citep{Bate09, Bate18}. Thus, multiple stellar systems with discs are expected to be common in star forming regions. This is also confirmed by surveys of Class 0 systems, as in \cite{Tobin+16}.

After the initial collapse of a molecular cloud core, the majority of the mass available to the forming stellar system is confined by angular momentum conservation in the disc and slowly accretes onto the stars \citep{Bonnell&Bate94}. 
The tidal torque between the central multiple system and the surrounding disc allows the exchange of angular momentum between the disc and the stellar system \citep{Lin&Papaloizou79, Goldreich&Tremaine80}. The gravitational torque exerted by the multiple system on the circum-multiple disc is thought to suppress the surface density in the surrounding of the stars. Indeed, a high enough angular momentum exchange between the system and the surrounding material is able to open a central cavity in the disc \citep{Artymowicz&Lubow96}. However, thanks to the asymmetries of the gravitational potential and to the three dimensional nature of the problem, accretion of gas onto the stars is not suppressed. Indeed, the stars of the system pull streamers of gas from the inner edge of the cavity. These streamers bridge the lower density region between the disc inner edge and the stellar system allowing the gas to flow towards the stars \citep{Artymowicz&Lubow96, Farris+14,Ragusa+16}. There, inner accretion discs around the single stars process the infalling gas that eventually is accreted.

An example is the well known GG Tauri A \citep{Keppler+20,Phuong+20}. GG Tau A is a triple \citep{DiFolco+14} stellar system surrounded by a circum-triple accretion disc. The stars carved a central cavity, where we observe streamers and filaments of gas. Another multiple system that shows cavities and nested discs separated by low density regions is the GW Orionis triple stellar system \citep{Kraus+20,Bi+20}. Other similar examples are the binary BHB 2007 \citep{Alves+19}, in which a complex structure of filaments supply gas from the circum-binary disc to the circum-stellar discs and L1448 IRS3B \citep{Reynolds+21}, that is really young multiple stellar system in formation.

In general, systems with more than two stars are unstable and their evolution eventually leads to the ejection of one body of the system \citep{Valtonen&Karttunen}. The only stable configurations observed are made of nested binary orbits and are called hierarchical configurations. For example, a hierarchical triple system is made of a binary orbited at distance by a third body. In order to preserve the stability of the system, the third body needs to orbit the binary at a distance of several time the binary semi-major axis (see \citet{MardlingAarseth01_triple_stability-crit} for a stability criterion).

How accreting mass from the circum-multiple disc distributes itself around the individual stars plays a key role in the star formation scenario. Indeed, both the evolution of the stellar system masses and the supply of gas around the stars to form inner discs (and possibly inner planets) strongly depend on the competition between the stars in having access to the gas stored in the disc. The study of these processes allows us also to link the properties of the observed evolved population of binaries to their initial conditions in which they initially born \citep{Bate00}.
The fraction of mass accreted by each star of the system depends on the system orbital parameters, in particular on the mass ratio , as shown by \citet{Farris+14}. In addition, other less studied system parameters play a role in the mass distribution among the system stars, for example the infalling gas temperature \citep{Young+15, Young&Clarke15} and the gas viscosity \citep{Duffell+20}.

This process, known as differential accretion, has been widely studied in binaries. Different numerical (SPH and grid) methods show that the secondary star of the binary should accrete most of the disc mass (e.g. in \cite{Bate&Bonnell97}, \cite{Farris+14}, and \cite{Young+15}). This is due to the lower relative velocity between the secondary and the disc material orbiting at the inner edge, and to the lower distance between the secondary orbit and the inner edge of the disc. However, there are exceptions to this general behaviour when the system orbit is very eccentric. Indeed, in this case, \cite{Dunhilletal15_binary_cavity-precession} showed that discs around binaries with mass ratio lower than unity can temporary accrete more mass onto the primary and \citet{Munoz&Lai16} showed that unitary mass ratio binaries can temporary break the symmetry expected in their accretion rates. Both these exceptions are due to the precession of the eccentric cavity carved by the stellar system.

As of today, however, little is known about differential accretion in hierarchical systems. In this paper we investigate to which extent the accretion trends of binary systems remain valid for hierarchical triples. In doing so, we propose a model to describe the deviations of the stellar accretion rates in triple systems. We also discuss the possibility to reveal unresolved hierarchical triple systems from their accretion rates and the difficulty in constraining the orbital parameters of the unresolved small binary.

The paper is organised as follows: in Sec.~\ref{sec:hydro} we describe the systems setup we considered and their initial conditions. In Sec.~\ref{sec:results} we present our results. We discuss the results of the simulations sets in Sec.~\ref{sec:discussion} and we give our conclusions in Sec.~\ref{sec:conclusions}.

\section{Hydrodynamical simulations}
\label{sec:hydro}

\begin{table}
\centering
\caption[]{Hydrodynamical simulations sets. All binary orbits (wider and smaller) are circular and coplanar with the disc. The semi-major axis of each wide orbit is $a_{\rm wide}=10$~au. All small binary have a unitary mass ratio $q_{\rm small}=1$.}
\label{tab:sims}
\begin{tabular}{cccc}
\hline
  Set 1 &  $\frac{a_{\rm small}}{a_{\rm wide}}$ & $q_{\rm wide}$ & split star \\
 \hline
	$b2$ &  - & 0.2  & -  \\	
    $b4$ &  - & 0.4  & -  \\
	$b65$ & - & 0.65 & -  \\
	 \hline
	$ts2$ & 0.1 & 0.2 & secondary  \\
	$ts4$ & 0.1 & 0.4 & secondary  \\
	$ts65$ & 0.1 & 0.65& secondary  \\
	$tp2$ & 0.1 & 0.2 & primary  \\	
	$tp4$ & 0.1 & 0.4 &primary  \\
	$tp65$ & 0.1 & 0.65& primary  \\
\hline
  Set 2 &  &  & \\
 \hline
 	$ts2a15$ & 0.15 & 0.2 & secondary  \\
	$ts2a05$ & 0.05 & 0.2 & secondary  \\
 \hline
	$ts4a18$ & 0.18 & 0.4 & secondary  \\
	$ts4a15$ & 0.15 & 0.4 & secondary  \\
	$ts4a05$ & 0.05 & 0.4 & secondary  \\
 \hline
	$ts65a20$ & 0.2   & 0.65 & secondary  \\
	$ts65a15$ & 0.15 & 0.65 & secondary  \\
	$ts65a05$ & 0.05 & 0.65 & secondary  \\
\hline
  Set 3 &  &  &   \\
 \hline
 	$tp2a20$ & 0.2 & 0.2 & primary  \\
	$tp2a05$ & 0.05 & 0.2 & primary  \\
 \hline
	$tp4a20$ & 0.2 & 0.4 & primary  \\
	$tp4a05$ & 0.05 & 0.4 & primary  \\
 \hline
	$tp65a20$ & 0.2   & 0.65 & primary  \\
	$tp65a05$ & 0.05 & 0.65 & primary  \\

\hline
\end{tabular}
\end{table}

We performed gas simulations of coplanar multiple systems embedded in an outer coplanar accretion disc using the 3D Smoothed Particle Hydrodynamics code {\sc Phantom} \citep{Price+18}. 

\begin{figure*}
    \centering
    \includegraphics[width=\textwidth]{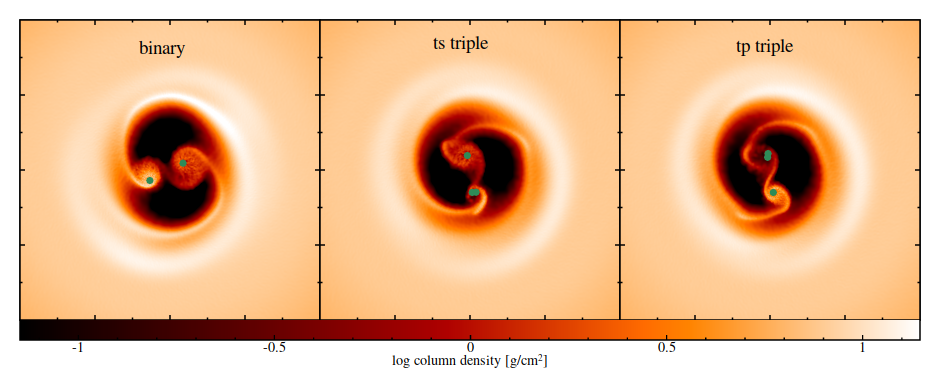}%
    \caption{Snapshots of three selected high resolution simulation of Set 1 showing the gas density in logarithmic scale. Green dots are the sink particles. All the three simulations have $q_{\rm wide} = 0.65$. On the first column there is the binary. The central column shows the triple system obtained by splitting the secondary star of the binary ($ts$ type). On the right column is shown the triple obtained by splitting the primary star of the binary ($tp$ type). All snapshots are taken at 55 wide binary orbits. 
    }%
    \label{fig:set1-splash}
\end{figure*}

We perform three sets of simulations. Set 1 consists of nine simulations made of three binary systems and six hierarchical triple systems. The three binaries have mass ratio $q_{\rm wide} = M_{\rm s}/M_{\rm p} = 0.2$, 0.4 and 0.65 respectively, where $M_{\rm s}$ is the mass of the lighter star and $M_{\rm p}$ the mass of the heavier one. Initially, the binaries are circular and have a semi-major axis $a_{\rm wide}=10$~au. From each binary we derived two hierarchical triple systems. The first kind of triples (labelled as $ts$) is built by splitting the secondary star of the binary, while the second kind of triples (labelled $tp$) is built by splitting the primary. In order to build the hierarchical triple systems, the binary stars are split into a circular binary with the same total mass of the split star, with a semi-major axis $a_{\rm small}=1$~au and a mass ratio $q_{\rm small}=1$\footnote{For more information about how we implemented in the {\sc Phantom} code the possibility to simulate hierarchical triple systems, see Appendix \ref{app-htphantom}.}. This set of simulations aims at understanding the effect of the mass ratio on the accretion trends. Table~\ref{tab:sims} contains the orbital configuration of each of these systems, and Fig.~\ref{fig:set1-splash} shows the gas surface density after 50 wide binary orbits for each simulation.

The simulations of Set 2 are devoted to explore the dependency of the triple system accretion rates on the small binary semi-major axis. In Set 2 we consider hierarchical triple systems where we split the secondary star. We start from the $ts2$, $ts4$ and $ts65$ simulations from Set 1 and we vary the small binary semi-major axis as reported in Table~\ref{tab:sims}.  

Finally, in Set 3 we focus on hierarchical triple systems where we split the primary star. We start from the $tp2$, $tp4$ and $tp65$ simulations from Set 1 and we vary the small binary semi-major axis as reported in Table~\ref{tab:sims}.

The total stellar mass of each binary and triple system is 3~$M_\odot$. Each system is surrounded by the same coplanar gas disc that initially extends from $R_{\rm in}=2\,a_{\rm wide}$ to $R_{\rm out}=10\,a_{\rm wide}$ with a mass equal to 0.03~$M_\odot$. The disc is modelled using $10^6$ SPH particles , resulting in a smoothing length about 0.1 times the disc scale height. The initial gas surface density profile is 
\begin{equation}
    \Sigma(R)=\Sigma_{\rm in}\left(\frac{R}{R_{\rm in}}\right)^{-p}\left(1-\sqrt{\frac{R_{\rm in}}{R}}\right),
\end{equation}
with $\Sigma_{\rm in}=69.4$ g cm$^{-2}$ and $p=1$. We assume a locally isothermal equation of state centered in the center of mass of the system. The sound speed profile follows
\begin{equation}
    c_{\rm s}(R)=c_{\rm s}(R_{\rm in})\left(\frac{R}{R_{\rm in}}\right)^{-q}, 
\end{equation}
with $q=0.25$. This results in a disc aspect ratio given by
\begin{equation}
    \frac{H}{R}=\frac{H_0}{R_0}\left(\frac{R}{R_{\rm in}}\right)^{(1/2-q)}.
\end{equation}
We set $H_0/R_0=0.1$ at $R=a_{\rm wide}$, as in \cite{Farris+14} and \cite{Young&Clarke15}.

Disc viscosity is implemented via the artificial viscosity method that is standard in SPH \citep{Lucy77, Gingold&Monaghan77}, which can be related to the \citet{Shakura&Sunyaev73} $\alpha$-viscosity as found by \citet{Lodato&Price10}. As differential accretion depends on viscosity, we set  $\alpha_{\rm SS}=0.1$, by setting $\alpha_{\rm AV}\approx9$, to match the values chosen by previous works, in particular by \cite{Farris+14} and \cite{Duffell+20}.

Stars are simulated as sink particles \citep{Price+18,Bate+95}. Sink particles are particles that interact only via gravity with other sink particles and SPH particles. They are evolved via a second-order Leapfrog integrator, as described in Section 2.8.5 of \citet{Price+18}. Sinks are allowed to accrete SPH particles and to store the accreted particles angular momentum and mass. The accretion of a gas particle can occur when it enters the accretion radius of a sink. To be accreted, the gas particle has to be gravitationally bound to the sink and its angular momentum has to be sufficiently low. In order to reliably resolve the accretion rates, the accretion radius of each sink is set to $0.1$~au. This radius is at most $\sim0.04$ times the wide binary secondary Roche lobe radius, depending on the binary mass ratio \citep{Eggleton83}.

All our simulations were evolved for 100 wide binary orbits, that correspond to half a viscous time-scale at the disc inner edge $R_{\rm in}\approx2a_{\rm wide}$, which can be expressed as \citep{Lynden-Bell&Pringle74, Hartmann+98}:
\begin{equation}
    \left.t_\nu\approx\frac{4}{9}\frac{R^2}{\nu}\right|_{R_{\rm in}},
\end{equation}
with $\nu=\alpha H c_{\rm s}$. We can write
\begin{equation}
    \nu = \alpha H^2 \Omega=\alpha\left(\frac{H}{R}\right)^2\Omega_{\rm wide}\sqrt{Ra_{\rm wide}^3},
    \label{eq:nu}
\end{equation}
where $\Omega$ is the Keplerian frequency and $\Omega_{\rm wide}$ is the binary orbital frequency.
Using Eq.~(\ref{eq:nu}) at $R=R_{\rm in}$, the viscous time in unit of binary orbits is
\begin{equation}
    \left.t_\nu\right|_{R_{\rm in}}\approx\frac{8}{9\sqrt{2}\pi}\frac{1}{\alpha(H/R)^2}\frac{2\pi}{\Omega_{\rm wide}}.
\end{equation}
With our choice of $\alpha$ and $H/R$ the viscous time is approximately $200$ binary orbits. We discuss the tests we made on longer integration time in Appendix \ref{app-tests}.

\begin{figure*}
\centering
\begin{subfigure}[h]{0.33\textwidth}
\centering
\includegraphics[width=\textwidth]{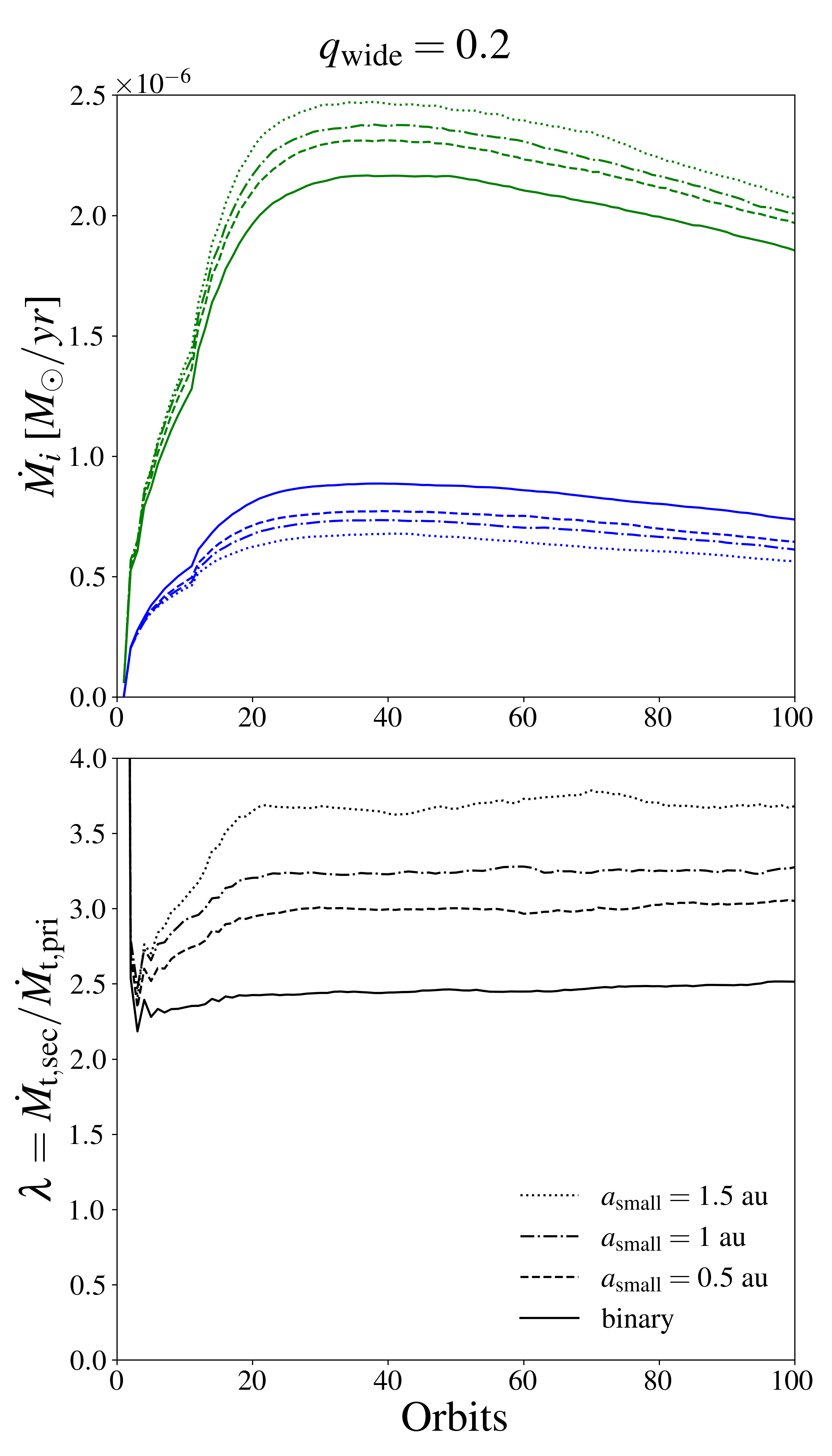} 
\end{subfigure}
\hfill
\begin{subfigure}[h]{0.33\textwidth}
\centering
\includegraphics[width=\textwidth]{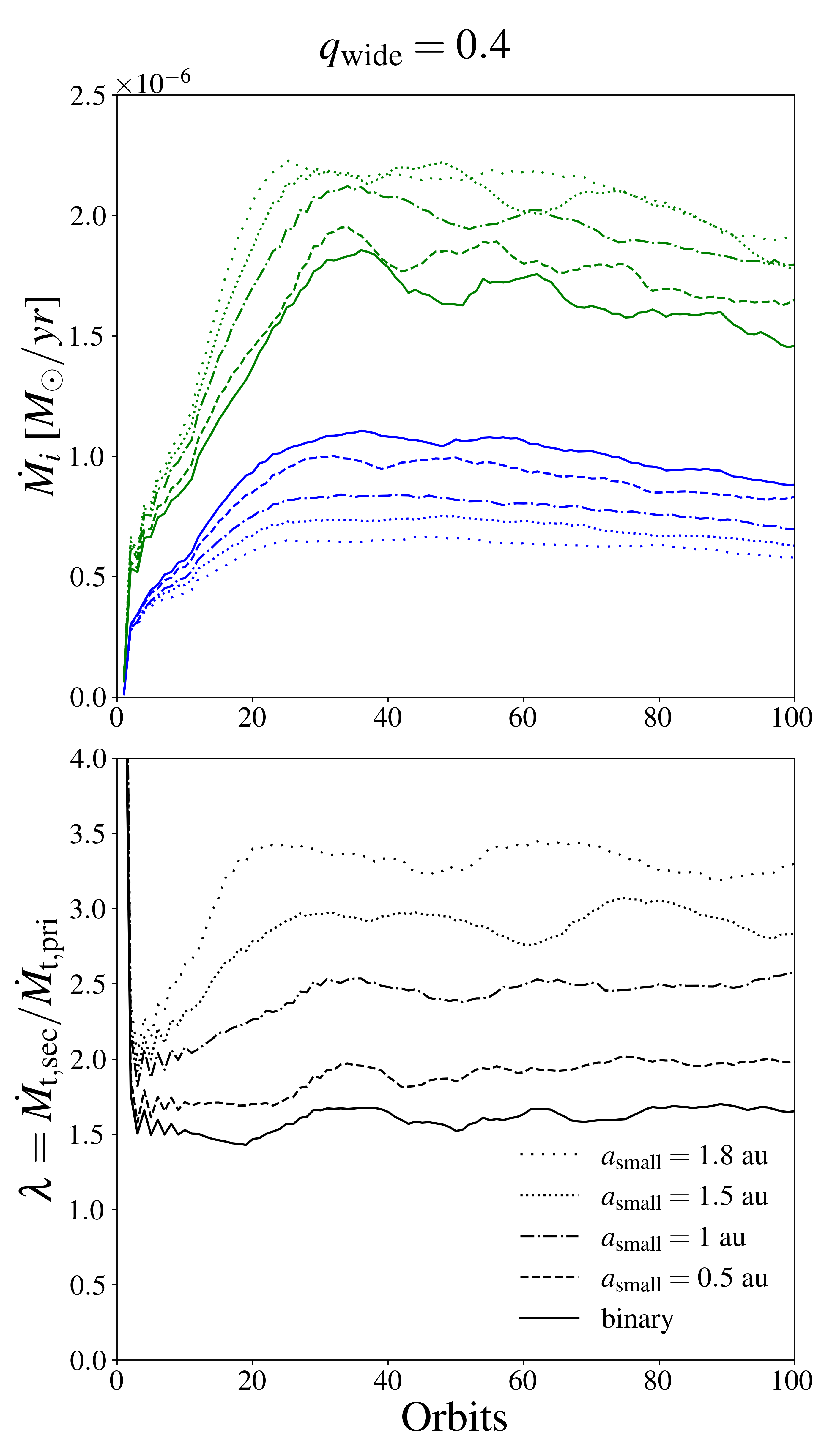} 
\end{subfigure}
\hfill
\begin{subfigure}[h]{0.33\textwidth}
\centering
\includegraphics[width=\textwidth]{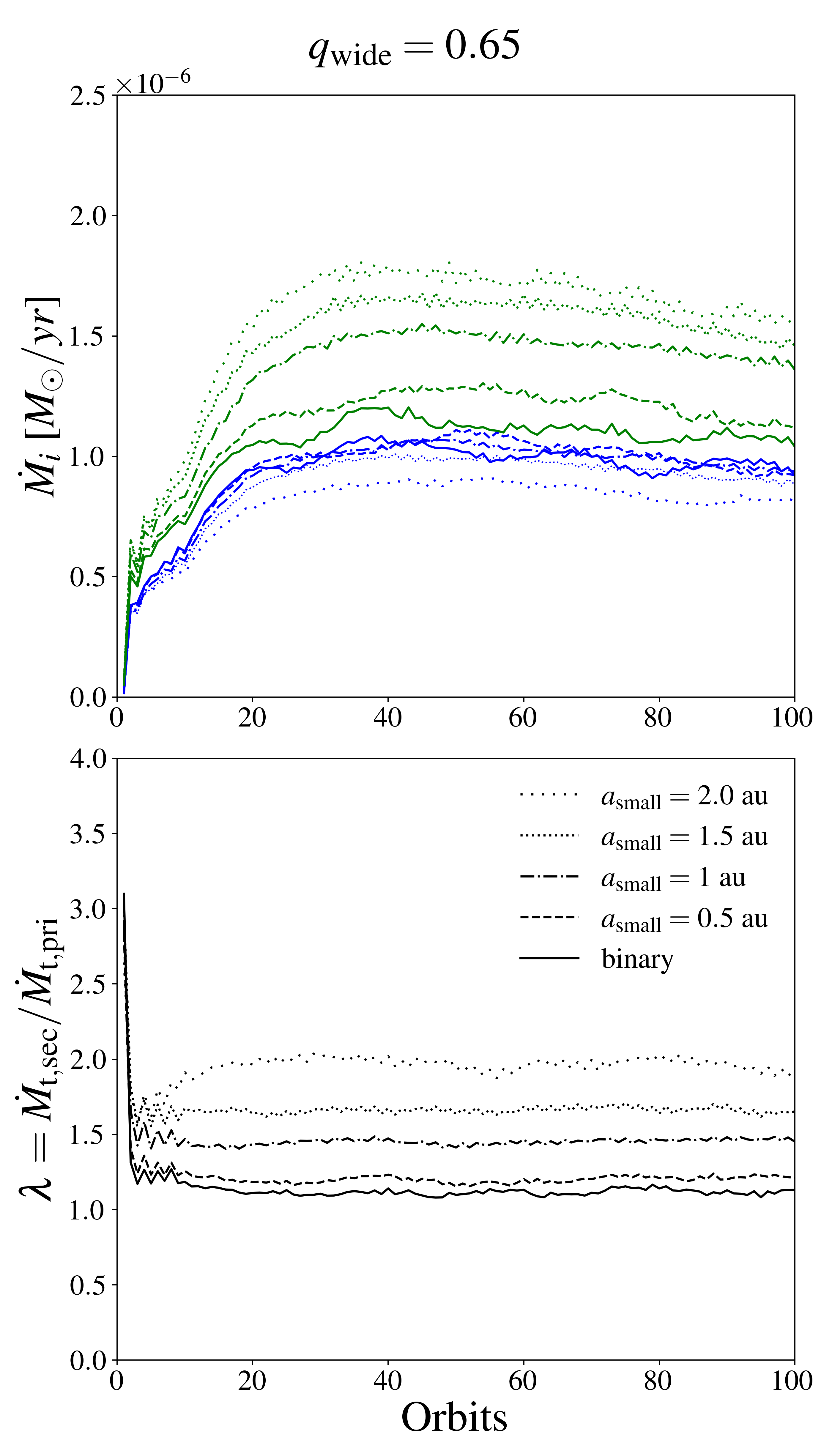} 
\end{subfigure}

 \caption{Moving averaged accretion rates and $\lambda=(\dot{M}_{\rm t,sec1}+\dot{M}_{\rm t,sec2})/{\dot{M}_{\rm t,pri}}$ factors measured in the $ts$ triple simulations (Set 2). For each wide orbit mass ratio $q_{\rm wide}$, on the upper panel are plotted the accretion rate of the secondary (green) and primary (blue) star. On the lower panel are plotted the ratio between the accretion rates. The solid line refers to the associated binary, from which the triples are generated. Each different line style refer to different small binary semi-major axis. The secondary accretion rate for triple system is the sum of the accretion rates of the small binary stars.}
\label{fig:ts-Mdot}
\end{figure*}

\begin{figure*}
\centering
\begin{subfigure}[h]{0.33\textwidth}
\centering
\includegraphics[width=\textwidth]{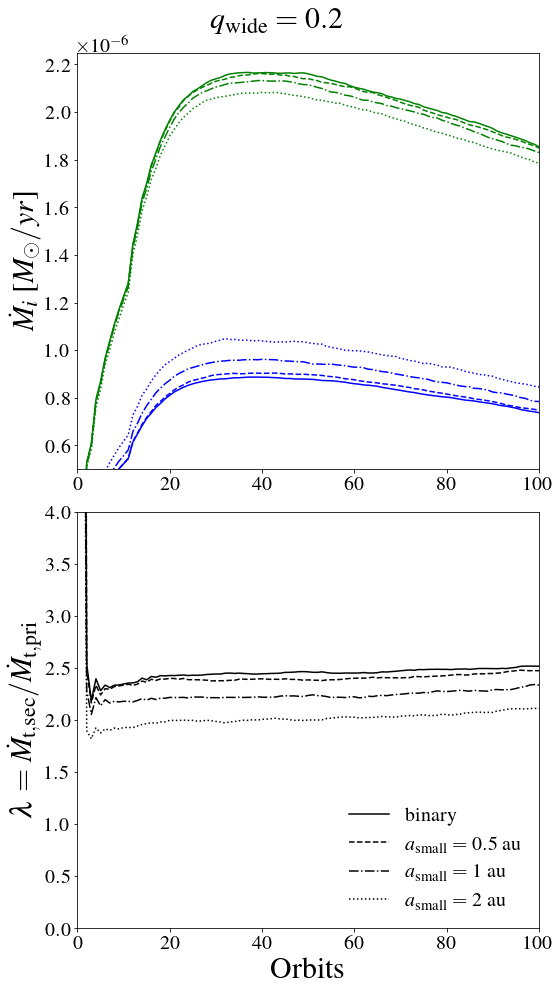} 
\end{subfigure}
\hfill
\begin{subfigure}[h]{0.33\textwidth}
\centering
\includegraphics[width=\textwidth]{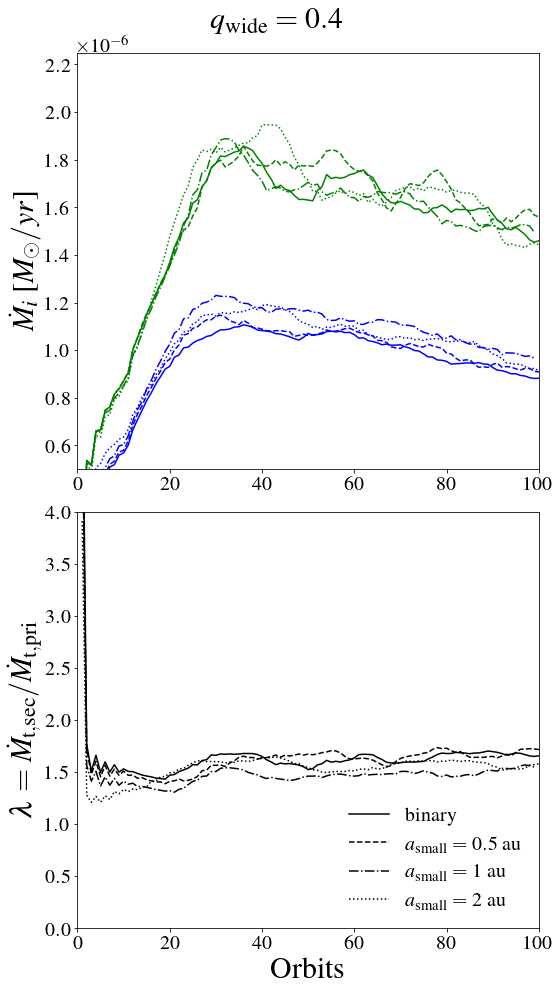} 
\end{subfigure}
\hfill
\begin{subfigure}[h]{0.33\textwidth}
\centering
\includegraphics[width=\textwidth]{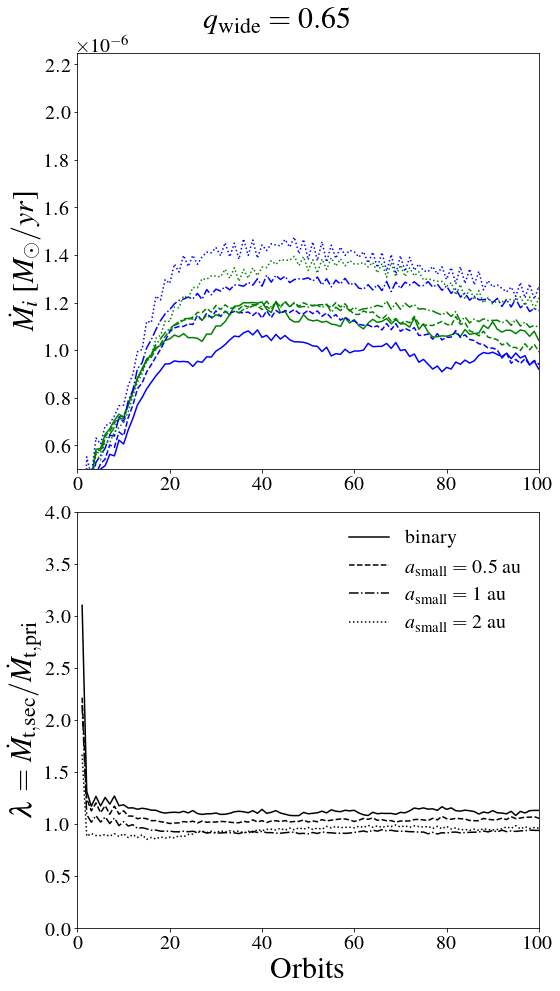} 
\end{subfigure}

\caption{Moving averaged accretion rates and $\lambda={\dot{M}_{\rm t,sec}}/({\dot{M}_{\rm t,pri1}+\dot{M}_{\rm t,pri2}})$ factors measured in the $tp$ triple simulations (Set 3). For each wide orbit mass ratio $q_{\rm wide}$, on the upper panel are plotted the accretion rate of the secondary (green) and primary (blue) star. On the lower panel are plotted the ratio between the accretion rates. The solid line refers to the associated binary, from which the triples are generated. Each different line style refer to different small binary semi-major axis. The primary accretion rate for triple system is the sum of the accretion rates of the small binary stars.}
\label{fig:tp-Mdot}
\end{figure*}

The main observable to be measured in this work is the accretion rate of each star during the simulation. 
The accretion rates of the simulations conducted in this work are shown in the first row of Fig.~\ref{fig:ts-Mdot} and Fig.~\ref{fig:tp-Mdot}.
We are not interested in the absolute value of the accretion rates but in their ratio. The ratio between the stellar accretion rates cancel out the decreasing trend shown in Figs.~\ref{fig:ts-Mdot} and \ref{fig:tp-Mdot} because the gas mass distributes with the same proportion between stars, in agreement with what found in \cite{Munoz+20}. In the second row of each mass ratio in Fig.~\ref{fig:ts-Mdot} and Fig.~\ref{fig:tp-Mdot} these ratios display a constant trend with an initial transient phase shorter than 20 orbits, showing that differential accretion quickly reaches a steady state.

The accretion rates measured in this work are reported for each orbit, as in Fig.~\ref{fig:ts-Mdot} and Fig.~\ref{fig:tp-Mdot}. To compute the accretion rates in the $n$-th orbit we integrate the $\dot{M}(t)$ over the orbital period $P$, thus
\begin{equation}
    \dot{M}_n=\int_{t_n}^{t_n+P} \frac{\dot{M}(t)}{P} {\rm d}t,
\end{equation}
where $t_n$ is the initial time of the $n$-th orbit.
We then averaged with a moving average over 11 orbits (i.e. ~ 4 orbits of the cavity inner edge), so that
\begin{equation}
    \langle\dot{M}_n\rangle=\frac{\sum_n^{n+10}\dot{M}_i}{11}.
    \label{eq:avemdot}
\end{equation}

The ratios between the stellar accretion rates with their errors (as in Fig.~\ref{fig:set1}) are computed discarding the initial transient orbits.

\section{Numerical results}
\label{sec:results}
\subsection{Binary systems differential accretion}
\label{sec:binaries}

\begin{figure}
    \centering
    \includegraphics[width=0.476\textwidth]{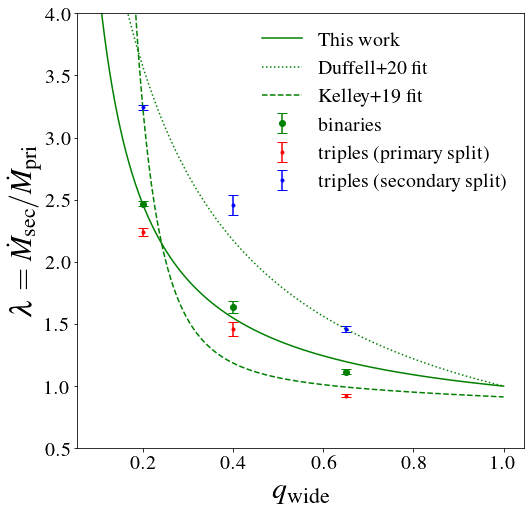}
    \caption{Set 1 simulations $\lambda=\dot{M}_{\rm sec}/\dot{M}_{\rm pri}$ values. Green dots are binary simulations. Red dots are triples obtained by splitting the primary star of the binary. Blue dots are triples obtained by splitting the secondary star. The solid green curve is the fit proposed in this work in Eq. (\ref{eq:binparam}). Dotted and dashed green curves are the \citet{Duffell+20} and \citet{Kelley+19} parametrisations, respectively.}
    \label{fig:set1}
\end{figure}

The ratio between the stellar accretion rates is the key observable in the binary systems differential accretion problem. Let us define this factor as:
\begin{equation}
    \lambda_{\rm b}=\frac{\dot{M}_{\rm b,sec}}{\dot{M}_{\rm b,pri}},
    \label{eq:lambda}
\end{equation}
where $\dot{M}_{\rm b,sec}$ and $\dot{M}_{\rm b,pri}$ are the moving averaged accretion rates of the secondary and primary star, respectively (defined in Eq.~(\ref{eq:avemdot})). The $\lambda_{\rm b}$ ratio measures how evenly the accreting mass distributes over the binary stars. If $\lambda_{\rm b}$ is larger than unity, this means more material is being accreted by the secondary. We simulate three binary systems (with $q_{\rm wide}=0.2, 0.4, 0.65$, simulations $b2, b4, b65$) in order to consistently compare the hierarchical triple simulations with their binary counterparts. Fig.~\ref{fig:set1} shows with green dots the $\lambda$ factors measured in our Set 1 of 3D SPH binary simulations.

The $\lambda_{\rm b}$ factor depends on the parameters of the system. In particular $\lambda_{\rm b}$ depends on the mass ratio of the binary, as pointed out by \cite{Farris+14}. In addition, $\lambda_{\rm b}$ depends also on the infalling gas properties (\citealt{Young+15}; \citealt{Young&Clarke15}). For a given mass ratio, warmer discs raise the primary accretion rate, pushing $\lambda_{\rm b}$ towards unity. This is due to the fact that warmer gas streamers have a wider range of trajectories to reach the primary star. In addition, warmer gas around the secondary star crosses the Roche lobe more easily, reaching the primary Roche lobe.
Last, \cite{Duffell+20} showed that $\lambda_{\rm b}$ depends also on gas viscosity. In particular, they found that for less viscous discs the value of $\lambda_{\rm b}$ tends towards unity.

Recently two parametrisations for $\lambda_{\rm b}(q)$ were proposed. The first one in \cite{Kelley+19} (hereafter K19 parametrisation), that updates the one proposed by \cite{Gerosa+15} and is built by fitting the \cite{Farris+14} binary simulation set. The second one in \cite{Duffell+20} (hereafter D20 parametrisation), who simulate binary accretion discs slowly modifying the binary mass ratio during the simulation in order to span $\lambda_{\rm b}(q)$ continuously. These works, based on different 2D grid numerical techniques, resulted in two different parametrisations (see green curves in Fig.~\ref{fig:set1}). In order to be able to compare our binary simulations with the D20 and K19 parametrisations, we used the same disc thickness and viscosity of previous works, even if higher than the typical protostellar disc viscosity \citep{Hartmann+98, Dullemond+18}.

As shown in Fig.~\ref{fig:set1}, our binary simulations are in fairly good agreement with the parametrisations proposed in the literature. In particular, we found the same accretions trends described by previous works. Indeed, the secondary star always accretes most of the mass. Moreover, the higher the binary mass ratio, the lower the value of $\lambda_{\rm b}$ (as expected). In addition, if we reduce the thickness of our disc, $\lambda_{\rm b}$ tends to the D20 parametrisation. The discrepancies can be due to the different numerical technique we used. In particular, our simulations are 3D as opposed to the 2D ones by \cite{Duffell+20} and \cite{Farris+14}, and the disc height profile could vary between simulations away from the inner cavity edge.

We fit our binary data points with the following one-parameter function:
\begin{equation}
   \lambda_{\rm b} = C + \frac{1-C}{q_{\rm wide}} \,\,\,,
   \label{eq:binparam}
\end{equation}
that accounts for the $q_{\rm wide}^{-1}$ dependency found in previous studies \citep{Gerosa+15, Duffell+20} and that approaches unity when $q_{\rm wide}$ approaches unity. Indeed, for symmetry reasons we expect a unitary mass ratio binary to evenly accrete mass onto the two binary stars. Our best fit for the $C$ parameter results in $C=0.63$. Contrary to K19 and D20 parametrisations, our formula is obtained from a set of 3D simulations and it is shown in Fig.~\ref{fig:set1} (green solid curve).

\subsection{Hierarchical triples differential accretion}
For a quantitative comparison with the $\lambda_{\rm b}$ factor measured in binaries, we introduce an analogous ratio for hierarchical triples: $\lambda_{\rm t}$. If the small binary of the triple system is lighter than the single body (i.e. in the $ts$ type triples) we define $\lambda_{\rm t}$ as the ratio between the the sum of the accretion rates of the small binary stars ($\dot{M}_{\rm t,sec1}+\dot{M}_{\rm t,sec2}$) over the accretion rate of the single star ($\dot{M}_{\rm t,pri}$):
\begin{equation}
    \lambda_{\rm t}=\frac{\dot{M}_{\rm t,sec1}+\dot{M}_{\rm t,sec2}}{\dot{M}_{\rm t,pri}}.
    \label{eq:lambdats}
\end{equation}

If instead the small binary is heavier than the single body (i.e. in the $tp$ triple case) we define $\lambda_{\rm t}$ as the ratio between the accretion rate of the single star ($\dot{M}_{\rm t,sec}$) over the sum of the accretion rates of the small binary stars ($\dot{M}_{\rm t,pri1}+\dot{M}_{\rm t,pri2}$):
\begin{equation}
    \lambda_{\rm t}=\frac{\dot{M}_{\rm t,sec}}{\dot{M}_{\rm t,pri1}+\dot{M}_{\rm t,pri2}}.
    \label{eq:lambdatp}
\end{equation}
In other words, the $\lambda_{\rm t}$ factor of a hierarchical triple system is defined considering the system as a binary in which the small binary is treated as a single body, with an accretion rate equal to the sum of the accretion rate of the small binary stars. 

Accordingly, we define $q_{\rm wide}$ (the mass ratio of the wide orbit) for the $ts$ and the $tp$ triples case. In the former case we define
\begin{equation}
    q_{\rm wide}=\frac{M_{\rm t,sec1}+M_{\rm t,sec2}}{M_{\rm t,pri}},
\end{equation}
where $M_{\rm t,sec1}$ and $M_{\rm t,sec2}$ are the mass of the small binary primary and secondary respectively, and $M_{\rm t,pri}$ is the mass of the single body. In the latter case we define
\begin{equation}
    q_{\rm wide}=\frac{M_{\rm t,sec}}{M_{\rm t,pri1}+M_{\rm t,pri2}},
\end{equation}
where $M_{\rm t,pri1}$ and $M_{\rm t,pri2}$ are the mass of the small binary primary and secondary respectively, and $M_{\rm t,sec}$ is the mass of the single body. In Fig.~\ref{fig:set1} we show, for each $q_{\rm wide}$ in Set 1, the $\lambda_{\rm t}$ factors of the triple simulations, along with the $\lambda_{\rm b}$ factor of their associated binaries. 

The differential accretion in hierarchical triple is set by the combination between the binary differential accretion and the effects induced by the presence of the small binary. Thus, in order to isolate the contributions of the small binary to differential accretion we compared each hierarchical triple simulations with their associated binary. 
In order to do this, we associated to each binary system discussed in Sec.~\ref{sec:binaries} (i.e. each $b$ simulation) two hierarchical triple systems, obtaining the nine simulations of Set 1 (see Tab.~\ref{tab:sims}). The two associated triples are built by substituting the primary or the secondary binary star with a small binary, obtaining, respectively, the $tp$ and $ts$ triple simulations (refer to Appendix \ref{app-htphantom} for the details on how we implemented this substitution in the {\sc Phantom} code). The triple obtained by splitting the secondary star can be viewed as a massive body orbited by a lighter binary ($ts$ type). If instead the primary is split, the system consists of a massive binary orbited by a third lighter body ($tp$ type). The substituting small binary is circular, has a mass ratio $q_{\rm small}=1$ and has a semi-major axis $a_{\rm small} = 0.1\, a_{\rm wide}$, where $a_{\rm wide}$ is the semi-major axis of the wide binary orbit. The mass of the substituting small binary is equal to the substituted star. With this process we built up the simulation Set~1, discussed in Sec.~\ref{sec:hydro}. 

As shown in Fig.~\ref{fig:set1}, in the parameter space region explored by this simulation set, $ts$ and $tp$ simulations raise their small binary accretion rate with respect to their single counterpart in the assocciated binary system. As a consequence, $ts$ simulations raise their $\lambda$ value while $tp$ simulations lower it. In addition, $ts$ triples shift $\lambda$ more than $tp$ triples. In Fig.~\ref{fig:ts-Mdot} and Fig.~\ref{fig:tp-Mdot} we report the accretion rates and $\lambda$ factors of every simulation of this work. The binary (solid curve) and the $a_{\rm small}=1$~au $ts$ and $tp$ simulations (dashed curves) show how the accretion rates of the single stars contribute in shifting the $\lambda$ factors. From the single accretion rates data, we see that the small binary always increases its accretion rate while the tertiary star lowers it. However, the split affects the single star only for lower mass ratios. Indeed, in the $q_{\rm wide}=0.65$ case the accretion rate of the single body does not change significantly. This implies that the total accretion rate of the $q=0.65$ triple systems is not conserved with respect to the binary case.
Given that the total accretion rate of the system is set by viscous accretion, it should not be different for different stellar systems surrounded by the same disc. Thus, we conclude that the $q_{\rm wide}=0.65$ systems have not reached steady-state. However, we show in Appendix \ref{app-tests} that 100 outer binary orbits are enough to measure $\lambda$ factors in reliable manner.

\subsection{Dependency on $a_{\rm small}$ and prescription for accretion rate deviations in triples}
\begin{figure}
    \centering
    \includegraphics[width=0.476\textwidth]{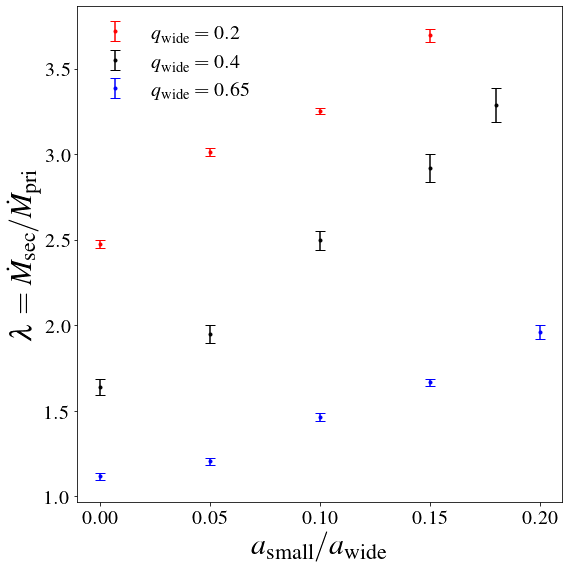}
    \caption{$\lambda=\dot{M}_{\rm sec}/\dot{M}_{\rm pri}$ factors for Set 2 triples. Different colours refer to different mass ratio of the wide orbit $q_{\rm wide}$. $a_{\rm small}/a_{\rm wide}=0$ points are the associated binary simulations.}
    \label{fig:tslambdas}
\end{figure}

\begin{figure}
    \centering
    \includegraphics[width=0.476\textwidth]{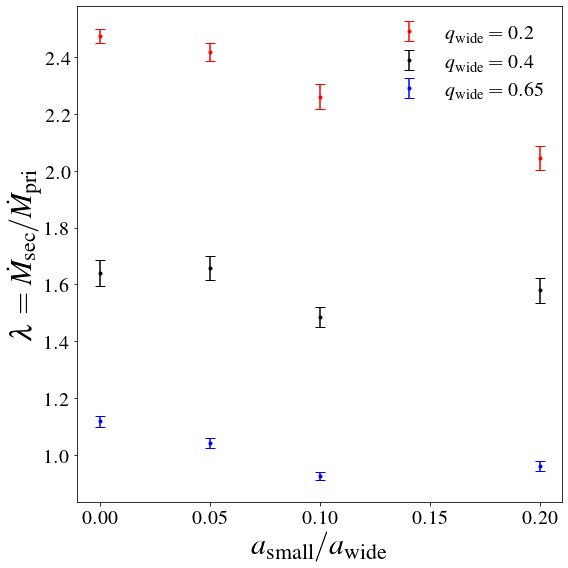}
    \caption{$\lambda=\dot{M}_{\rm sec}/\dot{M}_{\rm pri}$ factors for Set 3 triples. Different colours refer to different mass ratio of the wide orbit $q_{\rm wide}$. $a_{\rm small}/a_{\rm wide}=0$ points are the associated binary simulations.}
    \label{fig:tplambdas}
\end{figure}

How much accretion rates deviate from the binary case depends on the orbital configuration of the hierarchical triple, in particular on its small binary mass $M_{\rm b}$ and on its small binary semi-major axis $a_{\rm small}$. Indeed, Figs. \ref{fig:tslambdas} and \ref{fig:tplambdas} show how much the triple $\lambda_{\rm t}$ factor deviates from the associated binary $\lambda_{\rm b}$ factor as a function of $a_{\rm small}$ for Set~2 ($ts$ type triples, see Sec.~\ref{sec:hydro}) and Set~3 simulation ($tp$ type triples, see Sec.~\ref{sec:hydro}), respectively.

The change in the accretion rate of a multiple system star is linked to a variation of the net flux of mass in its Roche lobe. Indeed, in the steady state regime all the mass that enters the Roche lobe is eventually accreted by the star(s) inside the lobe. This implies that the mechanisms that modify the accretion rate of a body have to act on the scale of its Roche lobe. 
Two physics phenomena can be invoked in order to describe these deviations: the augmented geometrical cross-setion of the small binary and the angular momentum exchange between the small binary and the surrounding gas. 

On the one hand the small binary interacts with the surrounding gas through a geometrical cross-setion that is proportional to the area of the small binary orbit $A$,
\begin{equation}
    A\propto a_{\rm small}^2.
\end{equation}
Thus, we expect that the larger geometrical cross-setion of the small binary raises its accretion rate with respect to the corresponding single star in the associated binary system. 

On the other hand, we expect the small binary torque onto the surrounding gas to obstruct the accretion of material onto the small binary stars. In the impulse approximation \citep{Lin&Papaloizou79} we can estimate the torque exerted by the small binary onto the surrounding gas. If we suppose the small binary mass ratio $q_{\rm small}\ll1$, the density of tidal torque exerted by the small binary onto a fluid element at distance $p$ from the secondary star can be approximated by
\begin{equation}
    \tau=fq_{\rm small}^2\Omega_{\rm small}^2a_{\rm small}^2\left(\frac{a_{\rm small}}{p}\right)^4,
    \label{eq:impulse}
\end{equation}
where $\Omega_{\rm small}$ is the small binary frequency and $f$ is a dimensionless normalisation factor. Even if this approximation holds for low mass ratio binaries only, it gives us insights about how the specific torque scales with the small binary properties. We can assume that $p$ is approximately equal to distance between the inner binary stars and the small binary Roche lobe edge, thus $p\propto R_{\rm Roche}$. Hence, writing explicitly the binary frequency in Eq. (\ref{eq:impulse}) we obtain
\begin{equation}
    \tau \propto M_{\rm b}\frac{a_{\rm small}^3}{R_{\rm Roche}^4},
    \label{eq:torque-scale}
\end{equation}
where $M_{\rm b}$ is the small binary mass.

Taking into account the torque scaling and the geometrical cross-setion, we propose a parametrisation to describe the competition between these mechanisms in modifying the accretion rate of the triple small binary with respect to the corresponding single star in the associated binary system. With this prescription we also test the relative efficiency of different contributions to the deviations. Deviations are measured by means of the accretion ratio between the accretion rate of the small binary in the triple ($\dot{M}_{\rm t,sec}=\dot{M}_{\rm t,sec1}+\dot{M}_{\rm t,sec2}$) over the accretion rate of the corresponding binary star ($\dot{M}_{\rm b,sec}$). We then fit the accretion ratios with the following prescription:
\begin{equation}
\frac{\dot{M}_{\rm t,sec}}{\dot{M}_{\rm b,sec}}=1+\Gamma_\tau\frac{\left(a_{\rm small}/a_{\rm wide}\right)^3}{\left(R_{\rm Roche}/a_{\rm wide}\right)^{4}}+\Gamma_{\rm A}\left(\frac{a_{\rm small}}{a_{\rm wide}}\right)^{2},
\label{eq:prescr}
\end{equation}
where $\Gamma_\tau$ and $\Gamma_{\rm A}$ are parameters to be fitted and relate to the torque and the geometrical cross-setion, respectively. In Eq.~(\ref{eq:prescr}) we assume the geometrical term to scale with the small binary cross-setion ($\propto a_{\rm small}^2$), and the torque term to scale as in Eq.~(\ref{eq:torque-scale}) ($\propto a_{\rm small}^{3}/R_{\rm Roche}^4$). We thus expect that for small semi-major axis the cross-setion contribution to the accretion rate will dominate the accretion. Meanwhile, for wide semi-major axis the torque term will be more relevant. In addition, we expect that the torque parameter $\Gamma_\tau$ will be proportional to the small binary mass (as in Eq.~(\ref{eq:torque-scale})), while the geometrical cross-setion will depend only on the geometry of the orbits and not on the mass. However, the efficiency of each term depends on $\Gamma_\tau$ and $\Gamma_{\rm A}$. 

\subsubsection{Secondary split accretion ratios}
\label{sec:inner-axis}

In order to test the parametrisation for the accretion rate deviation due to the splitting, we simulate a second set of hierarchical triples. The aim of this set is to explore different regimes for the accretion ratio varying the semi-major axis of the small binary. 

We started from the $ts$ triples of Set 1 (with $a_{\rm small}=1$~au) and we simulated the same $ts$ hierarchical triple but with a semi-major axis of the small binary of $a_{\rm small}=0.5$ and $1.5$~au. We also exploited the wider stable range of semi-major axis for high $q_{\rm wide}$ hierarchical triples \citep{MardlingAarseth01_triple_stability-crit} in order to simulate even wider small binaries for $q_{\rm wide}=0.4$ (for which we also simulate $a_{\rm small}=1.8$~au) and $0.65$ systems (with $a_{\rm small}=2$~au). The accretion rates of this simulation set (called Set 2, see Sec.~\ref{sec:hydro}) are plotted in Fig.~\ref{fig:ts-Mdot} and their average $\lambda_{\rm t}$ factors are reported in Fig.~\ref{fig:tslambdas}.

\begin{figure}
    \centering
    \includegraphics[width=0.476\textwidth]{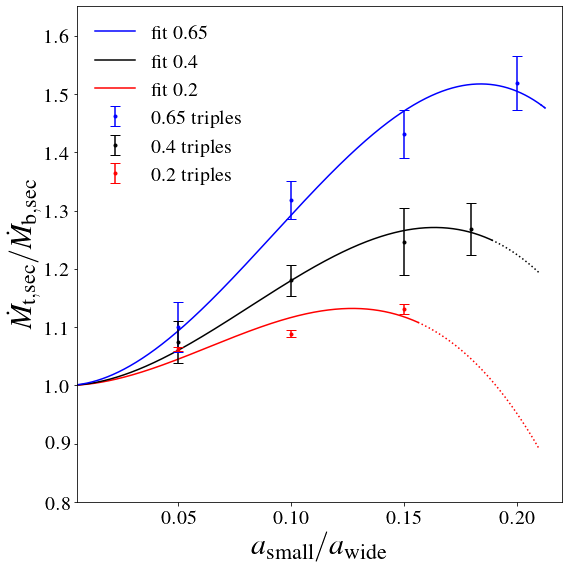}%
    \caption{Set 2 deviations in the small binary accretion rate as a function of the small binary semi-major axis. These simulations are obtained by splitting the secondary star of the three binaries of Set 1. The dots are the ratios between the accretion rate of the small binary ($\dot{M}_{\rm t,sec}$) and the accretion rate of the secondary star in the associated binary system ($\dot{M}_{\rm b,sec}$). The curves are obtained by fitting the three parameters of our prescription (Eq.~(\ref{eq:prescr})) for each mass ratio of the wide orbit (0.2, 0.4, 0.65 for red, black and blue curves respectively). The dotted part of each curve denotes the semi-major axes range for which the hierarchical triples are unstable.}%
    \label{fig:2-sma-fit}
\end{figure}

In order to apply the prescription in Eq.~(\ref{eq:prescr}), we compute the ratio between the accretion rate of the small binary ($\dot{M}_{\rm t,sec}$) over the accretion rate of the secondary star in the associated binary system ($\dot{M}_{\rm b,sec}$). When the repulsing effect of the binary torque contributes more than the geometrical cross-section to the accretion rate deviation, we expect this accretion ratio to be less than one. On the contrary, when the cross-section dominates, the ratio will be larger than one. Fig.~\ref{fig:2-sma-fit} shows $\dot{M}_{\rm t,sec}/\dot{M}_{\rm b,sec}$ varying the small binary semi-major axis.

Widening the inner binary, each $q_{\rm wide}$ data set in Fig. \ref{fig:2-sma-fit} follows a similar trend: a steep raise for smaller small binary semi-major axes, followed by lower deviations for larger small binary semi-major axes. Accretion ratios clearly depend on $q_{\rm wide}$, that in turn depends on the small binary mass. Indeed, for a given small binary semi-major axis, more massive small binaries systematically correspond to higher deviations in the accretion rate. Thus, the extent of the deviation depends on the split star mass. In the explored semi-major axes range, the accretion ratios are always higher than unity. Thus, the geometrical cross-section contribution to the deviation is greater than the torque contribution in each triple we considered.

\begin{figure}
    \centering
    \includegraphics[width=0.3\textwidth]{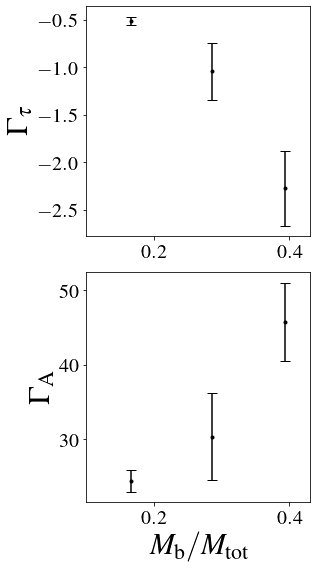}%
    \caption{Best fit $\Gamma$ parameters for our prescription (Eq.~(\ref{eq:prescr})) as a function of the small binary mass. $M_{\rm b}$ and $M_{\rm tot}$ are the mass of the inner binary and of the system, respectively.}%
    \label{fig:ts-params}
\end{figure}

In order to study how $\Gamma_\tau$ and $\Gamma_{\rm A}$ depend on the small binary mass, we separately fit the data point of each mass ratio (i.e. the blue, black and red points in Fig.~\ref{fig:2-sma-fit}) with the prescription in Eq.~(\ref{eq:prescr}). We thus obtain for each mass ratio the values of $\Gamma_\tau$ and $\Gamma_{\rm A}$ that best fit our data. These values are plotted in Fig.~\ref{fig:ts-params}. The parameter $\Gamma_\tau$ scales linearly with the small binary mass $M_{\rm b}$, as expected from Eq.~(\ref{eq:torque-scale}). Also $\Gamma_{\rm A}$ depends on the small binary mass, in contrast with what we expect from a purely geometric cross-section. The dependency of $\Gamma_{\rm A}$ on $M_{\rm b}$ can be due to a gravitational focusing effect. Indeed, without gravitational focusing, gas with an impact parameter higher than $a_{\rm small}$ skips the small binary geometric cross-section. On the contrary, in the gravitationally focused limit, gas with an impact parameter higher than $a_{\rm small}$ can enter the cross-section of the small binary. This is because the relative velocity between the gas and the binary is lower than the escape velocity from it. In this limit, more massive binaries have a larger effective cross-section, as noticed in the plot.

\subsubsection{Primary split accretion ratios}
\label{sec:primary}

We study how the accretion rate varies as a function of the orbital parameters of the triple when splitting the primary star. Primary and secondary stars are expected to accrete gas from the disc inner edge in different ways. On the one hand, the secondary has access to the gas stored in the disc mainly by pulling streamers directly from the inner edge. These streamers fill the secondary star Roche lobe (and to a lesser extent the cavity) with gas, allowing gas to fall onto the secondary star. On the other hand, the primary star pulls less massive streamers than the secondary, particularly for low mass ratios. Thus, the primary provides less gas directly from the disc, resulting in lower accretion rates. Another viable way for gas to reach the primary star is by means of the L1 point between the Roche lobes of the two binary stars. The more gas crosses the L1 point towards the primary, the higher its accretion rate at the expenses of the secondary star. As said, in hotter discs the accretion rate of the two stars are more even also thanks to this gas exchange \citep{Young+15}.

Raising the mass ratio $q_{\rm wide}$, these differences level out and the primary becomes more and more independent from the secondary in filling its Roche lobe with gas. As a result, the closer $q_{\rm wide}$ is to unity, the more the primary and secondary star accrete mass in a similar way. On the contrary, away from $q_{\rm wide}=1$ we expect the primary to be in a gas-poor environment, which prevents it from efficiently accreting mass. We expect these differences to result in different differential accretion deviations when splitting the primary star, rather than the secondary. Indeed, the split of the primary can either raise the mass that crosses L1 or raise the mass that falls onto the small binary from the inner edge. Thus, we tested the primary split configurations in Set 3.

In Set 3, we simulate a set of $tp$ hierarchical triples, based on the $tp$ simulations of Set 1 (with $a_{\rm small}=1$~au), varying the small binary semi-major axis ($a_{\rm small}=0.5, 2$~au, see Sec.~\ref{sec:hydro}). Fig.~\ref{fig:tp-Mdot} shows the accretion rates of Set 3 simulations and Fig.~\ref{fig:tplambdas} shows their average $\lambda_{\rm t}$ factors. 

The greater $\lambda$ factor deviations are observed in the $q_{\rm wide}=0.2$ systems (Fig.~\ref{fig:tplambdas}). The deviations are due to an enhanced flow through the L1 point, indeed, as shown in Fig.~\ref{fig:tp-Mdot}, the raise in the accretion rate of the small binary is at the expenses of the accretion rate of the third body. Fig.~\ref{fig:tplambdas} also shows that wider small binaries more easily capture mass from the third body Roche lobe, further reducing their $\lambda$ factor.

Hierarchical triples with $q_{\rm wide}=0.4$ and $0.65$ show smaller or no deviations due to the splitting. Indeed, in Fig.~\ref{fig:ts-Mdot} their $\lambda$ factors reduce up to 0.9 times the binary $\lambda$ factor. Thus, for higher mass ratios of the outer orbit the small binary is less efficient in stealing mass from the third body Roche lobe. In addition, and contrary to the $q_{\rm wide}=0.2$ case, wider small binary semi-major axes affect the deviations in $\lambda$ only modestly, as the impact of the geometrical cross-section is limited by the availability of mass in the surrounding of the small binary.

\begin{figure}
    \centering
    \includegraphics[width=0.476\textwidth]{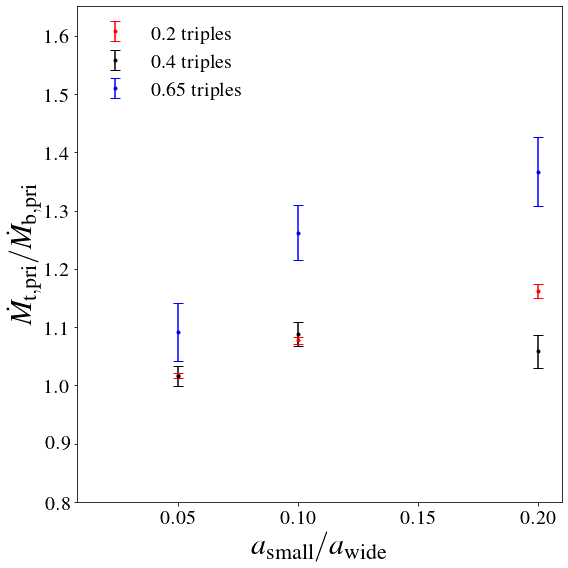}%
    \caption{Set 3 deviations in the small binary accretion rate as a function of the small binary semi-major axis. These simulation are obtained by splitting the primary star of the three binaries of Set 1. The dots are the ratio between the accretion rate of the small binary ($\dot{M}_{\rm t,pri}=\dot{M}_{\rm t,pri1}+\dot{M}_{\rm t,pri2}$) and the accretion rate of the primary star in the associated binary system ($\dot{M}_{\rm b,pri}$).}%
    \label{fig:1-sma}
\end{figure}

In light of this, the deviation observed in triples obtained by splitting the primary star cannot be captured by the effects described in Eq.~(\ref{eq:prescr}). In Fig.~\ref{fig:1-sma} we show the ratios between the accretion rate of the small binary over the accretion rate of its single counterpart in the associated binary system.
Only for high $q_{\rm wide}$ we recover the trend observed in Fig.~\ref{fig:2-sma-fit}, as the small binary starts to accrete more similarly to the secondary star of a binary, for which Eq.~(\ref{eq:prescr}) holds. 

It is important to notice that the accretion rate of primaries and $tp$ triples small binaries are not fully resolved (as discussed in Appendix \ref{app-tests}). Thus, the numerical results of this section have to be treated with caution. These results are still relevant as we report the \textit{relative} deviations due to the splitting measured for different choices of orbital parameters and we never rely on the absolute values that we measure in our simulations.

\section{Accretion in hierarchical triple systems}
\label{sec:discussion}
\subsection{Deviations of triple differential accretion from the associated binary system}
We have shown that hierarchical triples embedded in accretion discs have a peculiar way to distribute disc mass between the stars of the system. The ground state of differential accretion in hierarchical triples is based on the binary dynamics. Indeed, to a first approximation the wide orbit of the hierarchical triple mimics a binary system and it accretes mass in the same way, favouring the lighter body of the system. However, at smaller scales the influence of the triple system small binary has to be taken into account. The small binary-gas interaction changes the accretion rates of the three stars in relative terms (changing the proportion in which mass distributes among the stars, as shown in Fig.~\ref{fig:set1}).
Having a larger geometrical cross-setion, the small binary increases its accretion rate. This geometrical mechanism competes with the tendency of the small binary gravitational torque to repel the surrounding gas out of its Roche lobe. This competition gives the peculiar shape of the deviations in the accretion rate of the small binary as a function of its semi-major axis, as shown in Figs.~\ref{fig:2-sma-fit} and \ref{fig:ts-params}.

This mechanism generally shifts the triple system $\lambda_{\rm t}$ factor in favour of the small binary. However, how much $\lambda_{\rm t}$ deviates from the associated binary $\lambda_{\rm b}$ factor depends also on the mass ratio of the triple system wide orbit ($q_{\rm wide}$). Indeed, lower mass ratios show higher deviations from the associated binary, both when the small binary is lighter or heavier than the third body. Moreover, the triples obtained by splitting the secondary star of a binary result in higher deviations, compared to triples obtained by splitting the primary star (see Sec.~\ref{sec:primary}).

\subsection{Differential accretion in hierarchical triples}
The main consequence of differential accretion in binary systems is a tendency to equalise system masses. Indeed, as discussed in previous works \citep{Farris+14, Kelley+19, Duffell+20}, with enough mass at disposal the higher accretion rate of the secondary star pushes the mass ratio of the system towards unity. 

In this work we found that a hierarchical triple system in which the small binary is lighter than the third body raises the wide binary mass ratio $q_{\rm wide}$ more effectively than its associated binary system. Indeed, in the parameter space explored, the hierarchical triple $\lambda_{\rm t}$ factor (defined in Eq.~(\ref{eq:lambdats})) is higher than the $\lambda_{\rm b}$ factor of its associated binary (defined in Eq.~(\ref{eq:lambda})). 

For a quantitative comparison, the $q_{\rm wide}=0.65$ configuration (the one with the lowest $\lambda_{\rm t}$ among Set 1) has a $\lambda_{\rm t}\approx1.3\lambda_{\rm b}$, when splitting the secondary star of the associated binary. Even if we account only for the mass equalisation due to the triple mechanism (i.e. if we consider $\lambda_{\rm b}=1$) we found that the small binary accretion rate is $1.3$ times higher than the primary one. The binary differential accretion mechanism alone allows to obtain such a disequilibrium in the accretion rates only for $q_B$ lower than 0.35 or 0.75 for \citet{Kelley+19} and \citet{Duffell+20} parametrisations (respectively), as shown in Fig.~\ref{fig:set1}. For wider small binary semi-major axes or lower wide orbit mass ratios, the triple differential accretion mechanism is even more efficient. In addition, the disequilibrium between the stellar accretion rates in triples is at play even if $q_{\rm wide}\approx1$, where binary differential accretion is turned off.

We also remind that binary prescriptions strongly depend on the circum-binary gas properties and we do not know how they behave in actual protostellar discs conditions. On the contrary, we showed that the larger than unity $\lambda_{\rm t}/\lambda_{\rm b}$ ratio is due to the increased cross-setion of the small binary, which solely depends on the geometry of the orbits. Thus, we expect this ratio to be independent of the disc conditions. This difference is important because \citet{Duffell+20} showed that binary differential accretion is turned off in low viscosity regimes, where they found $\lambda \approx 1$ independently of the mass ratio. If this result is confirmed, we expect binary differential accretion to be turned off for low viscosity protostellar discs. But, if the larger than unity $\lambda_{\rm t}/\lambda_{\rm b}$ ratio is preserved (as we expect), the differential accretion in hierarchical triples constitutes the only viable mechanism to equalise the stellar masses.

\begin{figure}
    \centering
    \includegraphics[width=0.476\textwidth]{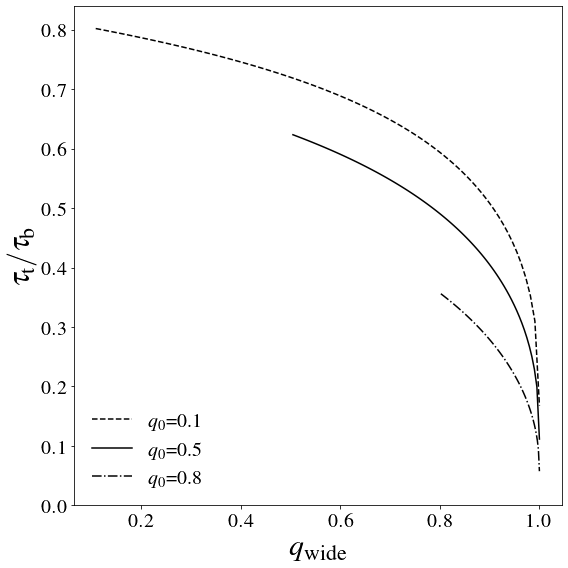}
    \caption{Ratio of the times needed in order to reach a certain $q_{\rm wide}$ by a triple ($\tau_{\rm t}$) and by a binary ($\tau_{\rm b}$). Different curves refer to different initial mass ratios $q_0$. Curves are obtained by solving Eq.~(\ref{eq:qdit}) with $\lambda(q_{\rm wide})=1$ for binaries and $\lambda(q_{\rm wide})=1.3$} for triples.
    \label{fig:eqtime}
\end{figure}

Assuming $\lambda_{\rm b}=1$, as in protostellar disc condition independently of $q$ \citep{Duffell+20} or as in more viscous discs around high $q$ systems, we can explicitly study the evolution of $q$ with time in the binary and in the hierarchical triple case. Under the approximation of a constant accretion rate (e.g. due to an infall that replenishes the outer part of the disc) we obtain the following differential equation for $q$:
\begin{equation}
\label{eq:qdit}
   \frac{{\rm d}q}{{\rm d}t_{\rm acc}} = \frac{q+1}{\lambda(q)+1}\left(\lambda(q)-q\right),
\end{equation}
where $t_{\rm acc}=(\dot{M}_{\rm tot}/M_{\rm tot})t$ is the time in unit of the mass doubling time of the system, with $\dot{M}_{\rm tot}$ and $M_{\rm tot}$ the total system accretion rate and mass respectively. Solving Eq.~(\ref{eq:qdit}) we obtain the times needed by a hierarchical triple and by a binary ($\tau_{\rm t}$ and $\tau_{\rm b}$, respectively) in order to reach a given wide orbit mass ratio $q_{\rm wide}$ starting from the same initial mass ratios $q_0$. Fig.~\ref{fig:eqtime} shows $\tau_{\rm t}/\tau_{\rm b}$ as a function of the final wide orbit mass ratio $q_{\rm wide}$ for three different initial mass ratio $q_0$. We assume $\lambda(q_{\rm wide})=1$ for binaries and $\lambda(q_{\rm wide})=1.3$ for triples, that is the triple $\lambda$ factor we expect from the $\lambda_{\rm t}/\lambda_{\rm b}$ ratio measured in our simulations, as discussed in the previous paragraphs. The equalising time for hierarchical triples (i.e. the time needed to reach $\lambda(q_{\rm wide})=1$) is nearly an order of magnitude lower than the binary equalising time, in particular for high $q_{\rm wide}$ systems.

A more subtle difference between the two mechanisms is the final equilibrium point of the wide orbit mass ratio $q_{\rm wide}$. Indeed, from Eq.~(\ref{eq:qdit}) we see that the equilibrium point for $q_{\rm wide}$ is at $q_{\rm wide}=\lambda$. Thus, because i) binary differential accretion prescriptions tends to unity for increasing $q_{\rm wide}$ and ii) symmetry reasons suggest that an equal mass binary has $\lambda_{\rm b}=1$, binaries stall at mass ratio $q_{\rm wide}=1$. Conversely, for a triple with a small binary heavier than the single body, we expect a $\lambda_{\rm t}$ lower than unity due to the raise in the small binary accretion rate. Thus, hierarchical triple systems stall at a mass ratio $q_{\rm wide}$ smaller than unity as well. The extent of this equilibrium shift depends on the specific orbital parameters of the system, which are responsible of the the shift in $\lambda_{\rm t}$. This work suggests an equilibrium point for triple systems of $q_{\rm wide}\approx0.9$. Indeed, $\lambda_{\rm t}/\lambda_{\rm b}$ measured in our triple systems with small binary heavier than the single star is approximately $0.9$. Thus, we expect that around $q_{\rm wide}=1$, where $\lambda_{\rm b}\approx1$ as well, $\lambda_{\rm t}=0.9$.

\subsection{Multiplicity signatures in differential accretion}

In principle, with a perfect knowledge of binary differential accretion and of its dependency on the binary mass ratio $q_{\rm wide}$, on the gas viscosity and temperature, we could be able to infer from an observed $\lambda_{b}$ in a binary system the presence of an unresolved small binary. Indeed, in case of binary accretion rates not in line with the binary theory, we could invert the relation proposed in Eq.~(\ref{eq:prescr}) in order to obtain the $a_{\rm small}$ of a possible unresolved small binary. However, up to now binary prescriptions do not take into account dependencies other than $q_{\rm wide}$\footnote{Although in \cite{Young&Clarke15} a trend related to temperature is suggested, further studies are needed to constrain an effective parametrisation.}. 
 
In addition, the disc conditions explored in this work (and in the main works on this topic in the literature) are halfway between the compact object accretion discs (that are thinner than the aspect ratio used) and protoplanetary discs (that are orders of magnitude less viscous), and thus do not represent either cases. 
Moreover, we limited our investigation to: i) circular wide and small binary orbits, ii) to equal masses small binaries and iii) to coplanar configurations. This allowed us to simplify the problem and observe the specific signatures of the geometrical and torque effects described in this work. 

We expect the mechanism proposed in this work to be at play in more complex configurations as well. However, its efficiency will be surely affected. This is mainly due to the dependency of both the accretion rate and the small binary torque on the orbital parameters of a given multiple system. In particular, the eccentricity and the mutual inclinations between the orbital planes and the discs are likely to play a major role since they can induced tilt oscillations and precession, which would translate into phase-dependent accretion rates along the orbit. A known example of a system where an highly eccentric binary shows phase-modulated accretion rates is discussed in \citet{Dunhilletal15_binary_cavity-precession}, where they show that for a limited amount of time it is possible for the primary to accrete more mass than the secondary.

Given these limitations, the only remaining case that at the moment could highlight an unresolved small binary in an accreting binary system is a system where a $\lambda<1$ is observed. In that case no binary configuration can reproduce this behaviour (except with an high eccentricity) and the only explanation that can be addressed to solve the puzzle should be a massive unresolved small binary, whose geometrical cross-setion counterbalances the tendency of binary differential accretion to favour the secondary single star. However, in our simulations even such configurations hardly push under 1 the ratio between the accretion rates, as can be seen in Fig.~\ref{fig:tp-Mdot} and discussed in Sec.~\ref{sec:primary}. Although in principle this should be possible for mass ratios greater than the ones explored in this work, the parameter space region where significant signatures of an hidden small binary could appear remains small. 

Thus, the goal to exploit the deviation of the observed $\lambda_{\rm t}$ in a triple system from the $\lambda_{\rm b}$ expected in its associated binary system is complicated by these additional dependencies and at the moment we cannot disentangle deviations due to different mechanisms, that change $\lambda$ without the need to invoke an higher multiplicity.

\section{Conclusions}
\label{sec:conclusions}
In this work we presented hydrodynamical simulations of discs in hierarchical triple systems. We focus on the accretion process from the circum-triple disc onto the individual stars of the system. In particular, we studied how the presence of the small binary affects the accretion rates of the individual stars. 

We performed a set of simulations in order to span different hierarchical triple system configurations using the SPH code {\sc Phantom}. We proposed a semi-analytical prescription (given by Eq.~(\ref{eq:prescr})) able to describe the data we obtained in our simulations.

Our main findings are the following:
\begin{enumerate}
\item Differential accretion in hierarchical triple systems can be explained by the interplay between two contrasting mechanisms: 1) the increased geometrical cross-setion between gas and small binary, and 2) its angular momentum exchange with it. These two mechanisms are superimposed on the binary differential accretion process. 
\item The small binary torque is too weak to counterbalance the increased accretion rate of the small binary due to the larger cross-section, except for very wide small binary semi-major axes ($a_{\rm small}$) that result in unstable hierarchical triples. Thus, in the vast majority of the stable hierarchical triple parameters space the small binary accretes more mass than if it would be a single star of the same mass. As a result, if the hierarchical triple small binary is heavier than the third body, the standard differential accretion scenario (whereby the secondary accretes more of the mass) is hampered. Reciprocally, if the small binary is lighter than the third body, the standard differential accretion scenario is enhanced.
\item Hierarchical triple systems with a small binary lighter than the single star equalise their masses nearly an order of magnitude quicker than binary systems. Conversely, in triples with a small binary heavier than the single star mass equalisation is slowed down. In contrast with binaries, the equilibrium mass ratio for triple systems is lower than 1.
\end{enumerate}

In conclusion, the mass ratio in accreting hierarchical triple stellar systems evolves differently compared to binaries. These differences, during the disc lifetime, are expected to produce characteristic mass ratio distributions, which could possibly be observed through ongoing and future surveys. Further observational data will help to test and further constrain the proposed accretion model for triple stellar systems. At any rate, the orbital parameters and initial masses play a crucial role in determining the final stellar mass ratios in high-order multiple stellar systems. 
	 
\section*{Acknowledgements}

The authors thank the referee for their constructive feedback and suggestions, which have significantly improved the original manuscript. This project has received funding from the European Union's Horizon 2020 research and innovation programme under the Marie Sk\l{}odowska-Curie grant agreements Nº 210021 and Nº 823823 (DUSTBUSTERS). SC thanks Cristiano Longarini and Pietro Curone for useful discussions. This work used {\sc Splash} \citep{Splash07} for SPH data visualisation. We also used the following Python tools and packages: {\sc NumPy} \citep{Numpy20}, {\sc Matplotlib} \citep{Matplotlib07} and {\sc Jupyter} \citep{Jupyter16}.

\section*{Data Availability}

The data underlying this article will be shared on reasonable request to the corresponding author. The code \textsc{Phantom} used in this work is publicly available at \url{https://github.com/danieljprice/phantom}.



\bibliographystyle{mnras}
\bibliography{triples} 




\appendix

\section{Numerical tests}
\label{app-tests}
\subsection{Accretion rate dependency on accretion prescription and spatial resolution}
The accretion rate onto the stars of each stellar system is the main observable measured in this work. In the following sections we detail the numerical tests we performed to check that the measured accretion rates are reliable. In the first section we discuss how the accretion rates depend on the accretion prescription we used in our simulations. In the second section we test if the measured accretion rates are fully resolved in our numerical simulations.

\subsubsection{Accretion prescription}
Given that we used sink particles, there is only one possible numerical choice when setting the simulation: the sink radius $R_{\rm sink}$. SPH particles inside a sphere of radius $0.8 \, R_{\rm sink}$ are automatically accreted onto the sink. The other particles inside a sphere of radius $R_{\rm sink}$ are accreted only if they are both gravitationally bound to the sink and have a sufficiently low angular momentum \citep{Price+18, Bate+95}. In all our simulations, we set all the accretion radii equal to $0.1$ au. This radius is roughly $4\%$ of the smallest Roche lobe radius around secondary stars in binaries (that is the smallest Roche lobe radius around small binaries in triples).

To test how the choice of sink radii affects stellar accretion, we ran two additional simulations doubling and halving all sink radii of our reference simulation (the $ts2$ triple in Set 1).
We integrated these two configurations for 100 outer orbit periods and we compared the accretion rates with the reference simulation. 

\begin{figure}
    \centering
    \includegraphics[width=0.476\textwidth]{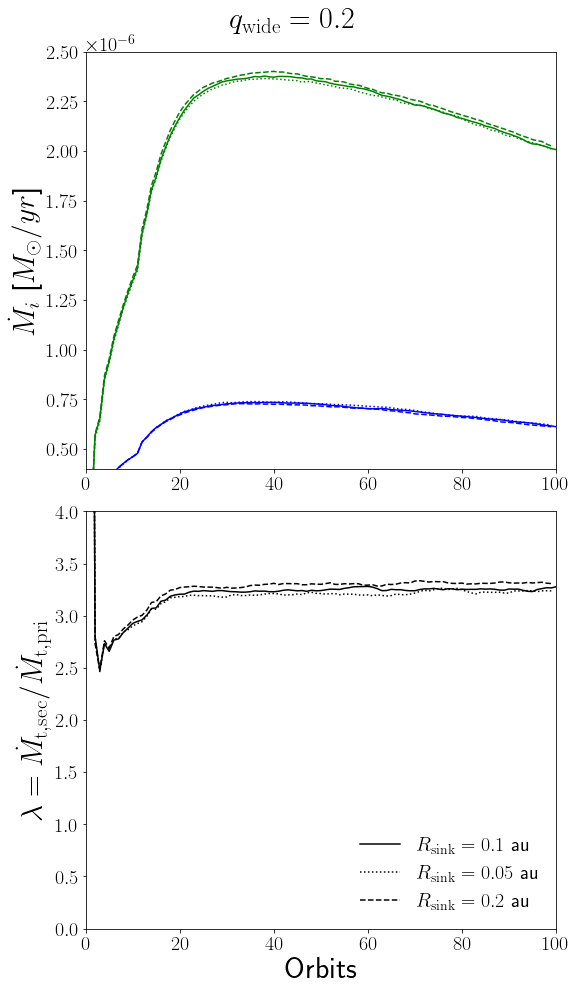}%
    \caption{Accretion rates and $\lambda=(\dot{M}_{\rm t,sec1}+\dot{M}_{\rm t,sec2})/{\dot{M}_{\rm t,pri}}$ factors measured in the triple simulations run to test the accretion prescription. On the upper panel are plotted the accretion rate of the secondary (green) and primary (blue) star. On the lower panel are plotted the ratio between the accretion rates. The solid line refers to the reference simulation $ts2$. Each different line style refer to different sink radii. The secondary accretion rate for triple system is the sum of the accretion rates of the small binary stars.}%
    \label{fig:sink-conv}
\end{figure}

Fig. \ref{fig:sink-conv} shows the accretion rates and $\lambda$ factors of test simulations, along with the reference one. We found that deviations from the reference simulation due to different sink radii are lower than 2\%. The deviations due to the splitting we measure in this work are at least one order of magnitude higher, particularly in triples with the small binary lighter than the third body (compare Fig. \ref{fig:sink-conv} with Fig. \ref{fig:ts-Mdot}). Since the accretion rates are not affected by the choice of the accretion radius, we conclude that our sink particles measure stellar accretion properly.

\subsubsection{Spatial resolution}
To model accretion onto the stars in a realistic way, we simulate the entire circum-triple disc. This choice limits our ability to carefully model the formation and evolution of circumstellar discs. The simulations presented in this paper barely resolve inner discs in the cavity of the circum-tripe disc. These discs do form within the cavity but with a limited spatial resolution. Indeed, the spatial resolution in the immediate surrounding of the stars is about 20\% of the Roche lobe radius (that is the spatial scale of the expected circum-stellar discs).

The net flux of mass through the Roche lobe boundary around each stars is the quantity that sets the individual accretion rates. The formation of discs inside the Roche lobes can introduce a delay during the disc formation phase. However, the mass that enters a given lobe will eventually fall onto the star. Indeed, mass cannot accumulate indefinitely in the Roche lobe. Here we wish to investigate whether the numerical resolution is high enough to ensure that the accretion rates onto the stars are well resolved.

\begin{figure}
    \centering
    \includegraphics[width=0.476\textwidth]{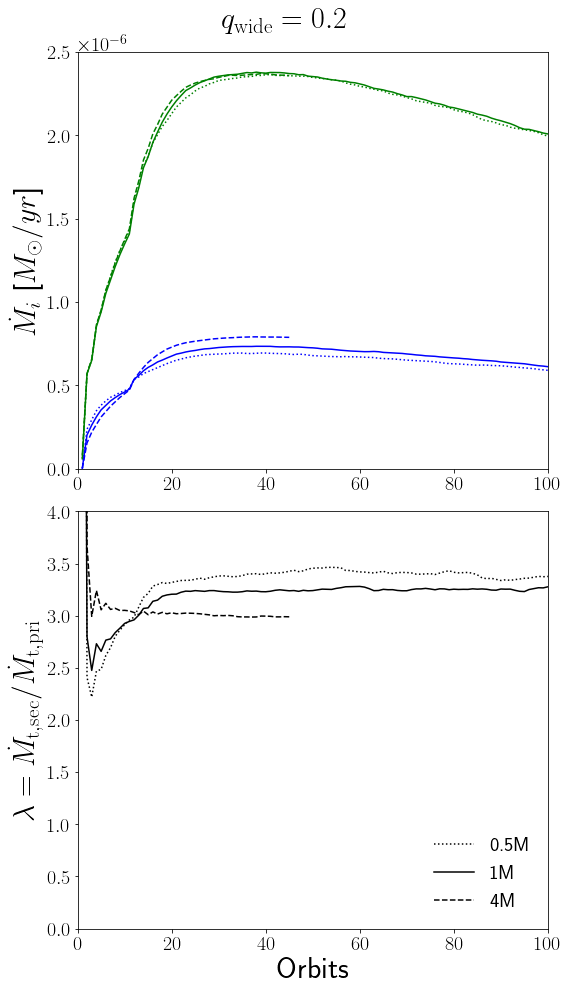}%
    \caption{Accretion rates and $\lambda=(\dot{M}_{\rm t,sec1}+\dot{M}_{\rm t,sec2})/{\dot{M}_{\rm t,pri}}$ factors measured in the triple simulations run to test the resolution. On the upper panel are plotted the accretion rate of the secondary (green) and primary (blue) star. On the lower panel are plotted the ratio between the accretion rates. The solid line refers to the reference simulation $ts2$. Dotted and dashed line refer to the lower and higher resolution simulations, respectively. The secondary accretion rate for triple system is the sum of the accretion rates of the small binary stars.}%
    \label{fig:res-conv}
\end{figure}

To test the resolution of our simulations we ran two additional simulation: we multiplied by 4 and divided by 2 the number of particles in the $ts2$ reference simulation, obtaining a higher resolution simulation of 4 millions particles and a lower resolution simulation with 500k particles. We ran the higher resolution simulation for 40 outer orbit periods and the lower resolution simulation for 100 orbits. Their accretion rates and $\lambda$ factors are shown in Fig. \ref{fig:res-conv}.

The simulations show that the accretion rate of secondaries are fully resolved: the simulations at higher and lower resolution exhibit the same accretion rates as the reference simulation. We note that the accretion rate of the primary grows with resolution. This implies that the circum-primary disc material is not fully resolved. Hence, we conclude that our results based on secondary splitting and the resulting accretion rate deviations are not affected by resolution issues. However, the measured $\lambda$ factors (both in binaries and in triples) are slightly overestimated, due to the underestimation of the primary accretion rate. The configuration we tested is the most affected by this issue, as it is the one with the lower primary accretion rate. Here, the $\lambda$ factor is overestimated by $\approx8\%$. It is worth highlighting that the ratio of $\lambda$ factors constitutes a more reliable quantity given that we are comparing binary and triple simulations \textit{at the same resolution}. However, our results about primary splitting (Sec. \ref{sec:primary}) have to be dealt with more caution. Our results are still relevant in the sense that -- instead of discussing individual accretion rates -- we report the relative deviations measured for different choices of orbital parameters.

\subsection{Integration time}
\begin{figure}
    \centering
    \includegraphics[width=0.476\textwidth]{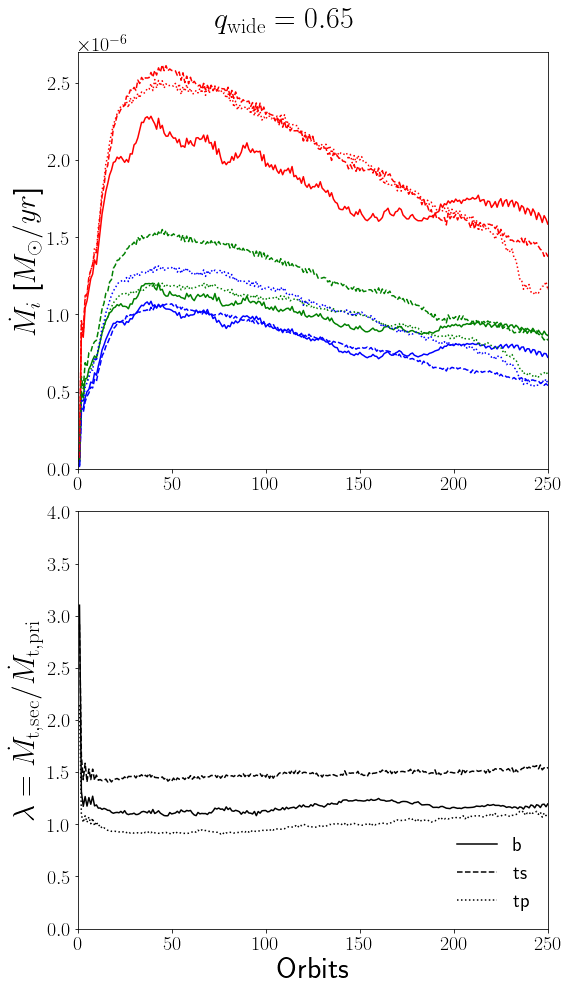}%
    \caption{Accretion rates and $\lambda=(\dot{M}_{\rm t,sec1}+\dot{M}_{\rm t,sec2})/{\dot{M}_{\rm t,pri}}$ factors measured in the triple simulations run to test longer integration times. On the upper panel are plotted the total accretion rate of the system (red) and the accretion rate of the secondary (green) and primary (blue) star. On the lower panel are plotted the ratio between the accretion rates (secondary over primary). The solid, dashed and dotted lines refer to the binary ($b65$), ts triple ($ts65$) and tp triple ($tp65$) simulations, respectively. The secondary (primary) accretion rate for ts (tp) triple system is the sum of the accretion rates of the small binary stars.}%
    \label{fig:int-time}
\end{figure}

Our simulations in sets 1, 2, and 3 last 100 outer orbit periods. As discussed in section \ref{sec:hydro}, this time span is half a viscous timescale at the inner edge of the disc. Thus, by the end of the simulation, the circum-triple disc has not reached steady-state. Given the number of simulations required to perform the analysis made in this work, modelling the entire disc viscous evolution is beyond our computational ability. Moreover, note that the configuration considered here, where the mass reservoir in the evolving disc is limited, will not actually settle into a steady state even for longer times. However, in this work we are interested in the way in which mass distributes from the circum-multiple disc over the stellar system stars. This is well measured by the ratio of the stellar accretion rate. In all the simulations we run, these ratios show an initial transient of less than 20 binary orbits -- regardless of the multiplicity and of the orbital parameters of the system, before settling down into quasi equilibrium. To further test this, we ran the three $q_{\rm wide}=0.65$ simulations in Set 1 for longer integration times. Fig. \ref{fig:int-time} shows the accretion rates and the $\lambda$ factors measured in these simulations. We find that, independently on the total accretion rate, mass divides between the sinks in the same way (i.e. we measure a constant $\lambda$ factor) until the gas smoothing length in the surrounding of the small binary exceeds the small binary semi-major axis. This happens around 250 orbits and it is a purely numerical effect caused by the loss of resolution around the small binary. Thus, even if the total accretion rate will evolve towards the steady-state, we expect the accretion rate ratios to remain constant. This allowed us to reliably measure the $\lambda$ factors even if our simulations have not reached a steady-state.

\section{Setting up hierarchical triple systems with phantom}
\label{app-htphantom}
The orbital arrangements of observed triple systems tend to be hierarchical, as different configurations are often unstable and have shorter life-times. A hierarchical triple (hereafter HT) system consists of a binary ($m_1$ and $m_2$) and a distant star ($m_3$) that orbits the center of mass of the inner binary. If the third body is sufficiently distant, an analytical perturbative approach is possible in order to compute the evolution of the system. In that case a first approximation of the inner and the outer orbit is the exact two-body orbit. Indeed, at each instant we can neglect the perturbations due to the \textit{triplicity} of the system and compute the orbital elements of the elliptical orbits that the three bodies would follow. These elements are called osculating elements. The set regarding the inner binary describes the orbit that the inner bodies would follow if the third body would instantaneously disappear. The set referring to the third body describes the orbit that it would follow if the inner binary was reduced to a single body with the total mass of the binary and in its center of mass.

In the case of a hierarchical triple system we can thus describe the instantaneous
state of the system again with 10 elements: the binary mass ratio
$q = m_2/m_1$, the triple mass ratio $Q = m_3/(m_1 + m_2)$, the semi-major axes
ratio, the two eccentricities, the two initial anomalies and the three Eulerian
angles to orient the orbits in respect to each other.

In order to simulate the stellar system configurations discussed in this work, we implemented in {\sc Phantom} the possibility to set as initial condition a hierarchical triple system embedded in a Keplerian circumtriple disc. Even if this work focused on coplanar orbit only, {\sc Phantom} is also able to set a misaligned hierarchical triple configuration. For future reference, we briefly describe the way in which hierarchical triples are initialised in the code.

The initial position and velocity of the two binary bodies are computed by means of the Thiele-Innes elements \citep{Binnendijk60}. Thiele-Innes elements are computed in terms of the Campbell elements through the following relations:
\begin{equation}
\begin{split}
\mathbf{P} = &(\cos\omega\cos\Omega - \sin\omega\cos i\sin\Omega, \\
 &\cos\omega\sin\Omega + \sin\omega\cos i\cos\Omega,\\
 &\sin\omega\sin i) \,\,\,, 
\end{split}
\label{eq:P}
\end{equation}
\begin{equation}
\begin{split}
\mathbf{Q} = &(-\sin\omega\cos\Omega - \cos\omega\cos i\sin\Omega, \\
 &-\sin\omega\sin\Omega + \cos\omega\cos i\cos\Omega,\\
 &\cos\omega\sin i) \,\,\,,
\end{split}
\label{eq:Q}
\end{equation}
\begin{equation}
    A = \cos{E}-e \,\,\, ,
\label{eq:A}    
\end{equation}
\begin{equation}
    B = \sqrt{1-e^2}\sin{E} \,\,\, ,
\label{eq:B}    
\end{equation}
where $\omega$, $\Omega$, $i$, $a$ and $e$ are the argument of the pericenter, the argument of the ascending node, the inclination, the semi-major axis and the eccentricity of the binary orbit, respectively, and $E$ is the eccentric anomaly. With Eq. (\ref{eq:P})-(\ref{eq:B}) we can compute the rectangular coordinates and velocities of a given initial condition as:
\begin{equation}
    (x,y,z) = a(A\mathbf{P}+B\mathbf{Q}),
\end{equation}
\begin{equation}
    (v_{\rm x},v_{\rm y},v_{\rm z}) = -a\dot{E}(\sin E\mathbf{P}-\sqrt{1-e^2}\cos E\mathbf{Q}),
\end{equation}
where $\dot{E}$ is the time derivative of the eccentric anomaly. The eccentric anomaly E and its derivative $\dot{E}$ are computed from the true anomaly of the orbit.

The hierarchical triple system initial condition is built by means of the osculating elements of the outer and of the inner binary orbit. Firstly the code builds a binary with the orbital parameters of the outer orbit. Then, one of the sinks of the outer binary is substituted with an inner binary of the same mass of the substituted sink. The center of mass of the inner binary follows the orbit of the substituted sink. The subroutine devoted to this task can be called as much times as needed, in order to build a generic hierarchical system, even with more than three stars. After the initial setup, the N-body dynamics of the system is solved as described in Section 2.8.5 in \cite{Price+18}.

Additional information and techincal details can be found in the online {\sc Phantom} documentation: \url{https://phantomsph.readthedocs.io/en/latest/hierarchicalsystems.html}


\bsp	
\label{lastpage}
\end{document}